# PHASE-SPACE ANALYSIS USING TOMOGRAPHY FOR THE MUON G-2 EXPERIMENT AT FERMILAB *

D. Stratakis[†], Fermi National Accelerator Laboratory, Batavia IL, USA

*Abstract*

In the next decade the Fermilab Muon Campus will host two world class experiments dedicated to the search for signals of new physics. The Muon g-2 experiment will determine with unprecedented precision the anomalous magnetic moment of the muon. The Mu2e experiment will improve by four orders of magnitude the sensitivity on the search for the as-yet unobserved Charged Lepton Flavor Violation process of a neutrinoless conversion of a muon to an electron. Maintaining and preserving a high density of particles in phase-space is an important requirement for both experiments. This paper presents a new experimental method for mapping the transverse phase space of a particle beam based on tomographic principles. We simulate our technique using a GEANT4 based tracking code, to ascertain accuracy of the reconstruction. Then we apply the technique to a series of proof-of-principle simulation tests to study injection and transport of muon beams for the Fermilab Muon Campus.

## INTRODUCTION

A common challenge for accelerator systems is to maintain beam quality and brightness over the usually long distance from the source to the target. In order to do so, knowledge of the beam distribution in both configuration and velocity space along the beam line is needed. However, in many occasions measurement of the velocity distribution can be difficult.

Here we present a simple and portable tomographic method to map the beam phase space, which can be used in the majority of accelerators. The tomographic reconstruction process has first been compared with results from simulations using the tracking code G4Beamline. Results show excellent agreement. Special emphasis is given to the Fermilab Muon Campus [1] beamlines where our phase space tomography diagnostic is used to optimize injection to the storage ring of the Muon g-2 Experiment [2].

## PHASE-SPACE TOMOGRAPHY

In order to analyze and understand the detailed behavior of the beam transport knowledge of the phase distribution is needed. Tomographic methods have shown in the past to recover high quality phase spaces without making a priori assumption of the initial conditions. Computerized tomography is well known in the medical community and was originally developed to process x-ray images. A Norwegian physicist Abel (1826) first formulated the concept of tomography for an object with axisymmetric geometry. Nearly 100 years later, an Austrian mathematician Radon (1917) developed a theorem extending the idea to arbitrarily shaped objects; it stated that an object in an n-dimensional space can be recovered from a sufficient number of projections on to (n−1)-dimensional space [3]. The principle of tomography is illustrated in Fig. 1.

In beam physics, we can map the phase-space using information taken from the distribution of spatial density at the same point. A simple scaling equation relates the spatial beam projections to the Radon transform of the transverse phase space, as demonstrated in the 1970s by Fraser et al. [4]. Specifically, the authors imaged the beam at different positions along the beam line and then reconstructed the phase-space distribution using tomographic computer programs.

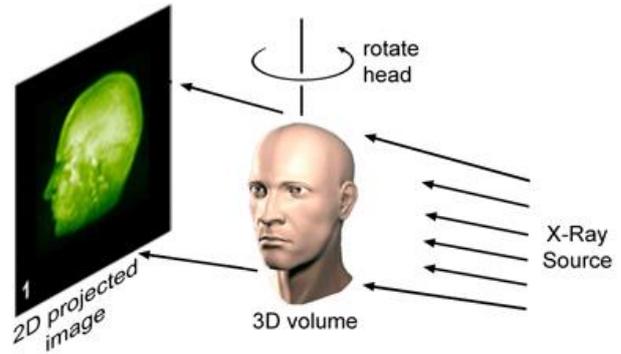

Figure 1: Principle of phase-space tomography as used in the medical community.

Phase-space tomography was implemented with greater accuracy by the study of McKee et al. [5] wherein they combined the ideas of tomography with quadrupole scanning to recover density information in phase space. Quadrupole scan techniques allowed the collection of multiple projections without intercepting the beam distribution. The main idea is that a simple variation of the magnet strength rotates the phase space distribution. By appropriate scaling, projections in configuration space are related to projections in phase space. Both scaling factor and angle of the projection are calculated from the beam transport matrix and detailed equations are reported elsewhere [6]. Since then, several authors adopted a similar approach. Examples of how projections in phase space are collected are demonstrated in Fig. 2.

## APPLICATION TO THE MUON CAMPUS

Protons with 8 GeV kinetic energy are transported via the M1 beamline to an Inconel target. Secondary beam from the target will be collected using a lithium lens, and positively-charged particles with a momentum of 3.1 GeV/c (± 10%) will be selected using a bending magnet. Secondary beam leaving the target station will travel through the M2 and M3 lines which are designed to capture

___

muons with momentum 3.1 GeV/c. The beam will then be injected into the Delivery Ring (DR). After several revolutions around the DR, essentially all pions will decay into muons, and the muons will separate in time from the heavier protons. A kicker will then be used to remove the protons, and the muon beam will be extracted into the M4 line, and finally into the M5 beamline which terminates just upstream of the entrance of the Muon g-2 Experiment storage ring. At the end of the M5 four quadrupole magnets are used to steer the beam towards the storage ring. These magnets are flexible and can be adjusted to the optimum conditions. The set-up is illustrated in Fig. 3. More details on the Muon Campus beamlines can be found in Ref. 2

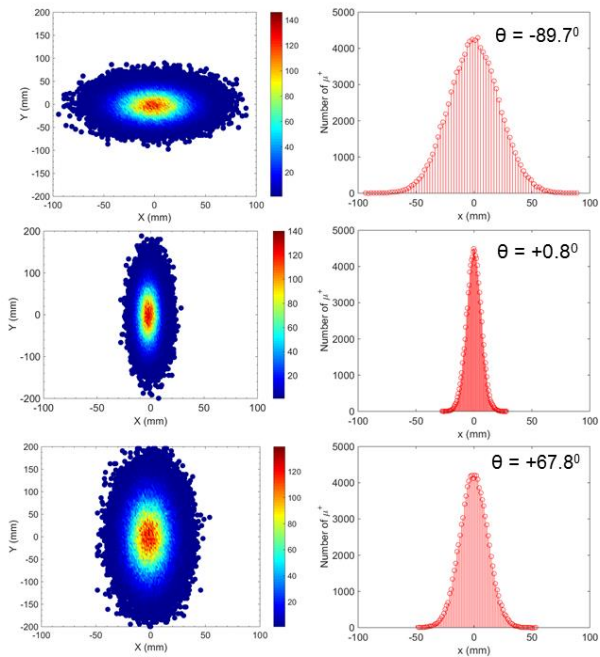

Figure 2: Left side shows the real spatial (x,y) beam photos and the right side shows the integrated x profiles for different focusing scenarios. The angles are calculated from the transport matrix by following the procedure described in Ref. 6.

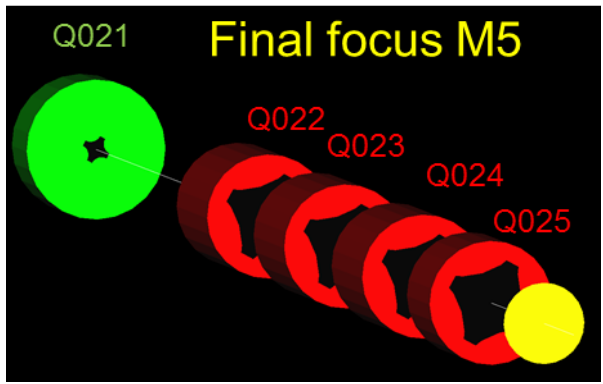

Figure 3: Schematic illustration of the final focus area at the end of M5. The last four magnets are powered independently and therefore are suitable for quad scans.

Beam monitoring along the Muon Campus can be divided into four different zones, each with different instrumentation schemes. High-intensity proton beam will be monitored with toroids, beam position monitors (BPMs) and beam loss monitors (BLMs). Low-intensity secondary and proton-only secondary beam will be monitored with ion chambers, BLMs and secondary emission monitors (SEMs). Muon-only beam will be monitored with ion chambers and proportional wire chambers (PWCs).

In particular, beam profiles in the upstream M4 and M5 beamlines will be measured using PWCs. PWCs are sensitive, since they have the capability to measure beam intensities down to the $10^3$ particle range. When mounted inside refurbished Switchyard bayonet vacuum cans, the PWCs can be pulled out of the beam path when not in use. This eliminates the need for permanent vacuum windows and vacuum bypasses. The PWC has two planes of signal wires, one plane for horizontal and one for vertical. There are 48 signal wires in each plane which are 10 μm diameter gold-plated tungsten and can be configured with either 1 mm or 2 mm spacing. The wire planes are sandwiched between Aluminum high-voltage bias foils where negative voltage is applied. Two end plates hold the entire assembly together. The PWC assembly is filled with an 80% Argon and 20% Carbon Dioxide gas mixture.

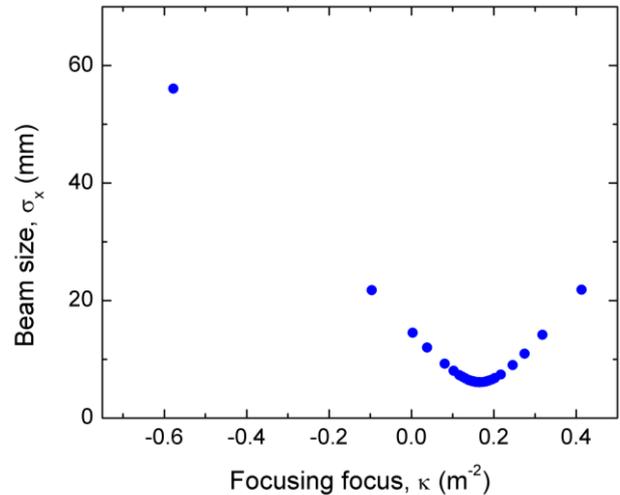

Figure 4: Beam size at the simulated profile monitor just downstream of Q025. The corresponding Q024 focusing strengths are shown in the horizontal axis.

To verify our proposed diagnostic we model a quadrupole scan and follow the process described in Ref. 6 in order to tomographically reconstruct the beam phase space. To accomplish this, we vary the strength of quadrupole magnet Q024 and collect beam profiles just downstream of quadrupole Q025. The set-up is illustrated in more detail in Fig. 3. Figure 4 shows the corresponding beam size on the screen versus the strength of the magnet Q024. The tomography reconstructed phase space is compared to the phase space generated directly by the simulation. The beam propagation was simulated using G4Beamline [7]. Here it is important to emphasize that the phase space generated by

G4Beamline does not make the assumptions that tomography does (tomography assumes constant emittance, ignores magnetic apertures, assumes no decays of muons) and therefore can be used as a prototype to establish the quality of our tomography method. For the simulation, we use assume a Gaussian muon beam similar to the one expected during the operation of the Muon g-2 Experiment. We initialize our simulation upstream of Q021 (see Fig. 3).

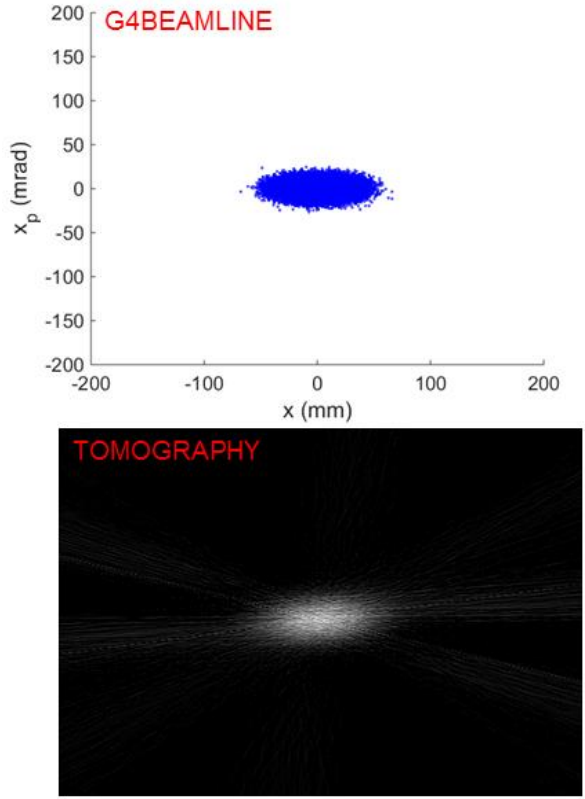

Figure 5: Horizontal phase-space distribution upstream of quadrupole Q021 created directly by G4Beamline and by tomography (bottom). For tomography we use 29 projections.

Figure 5 shows the phase space upstream of Q021. Top row shows the G4Beamline generated one while the bottom shows the one reconstructed with our tomographic method. In order to quantify the agreement, beyond the visual inspection, we are able to calculate from the reconstructed and simulated distributions the effective emittances and found values that are within 5% for the two cases. Figure 6 shows the phase-spaces in the same location but in the vertical direction. Similar to the horizontal case the agreement in emittances is within 5%.

In the upcoming months and during commissioning of the Fermilab Muon Campus, an experimental program to measure the beam phase-space with tomographic techniques will be established. Downstream of Q025 a PWC monitor is installed and will be capable to provide the required beam profiles for reconstruction. Figure 7 shows an example of such a profile. If successful, the diagnostic can prove very useful in optimizing the performance of the Muon g-2 Experiment. In particular, it will provide important information about the beam conditions at injection so that they can used as initial conditions for the subsequent simulations of the storage ring.

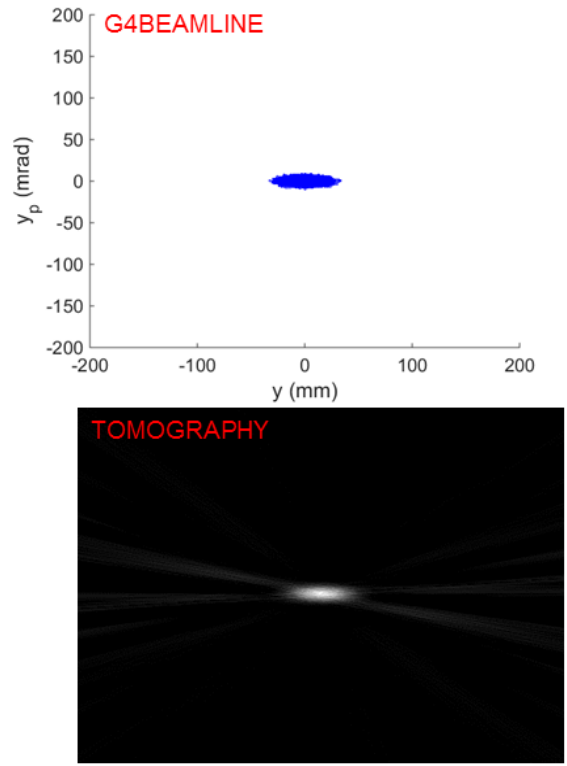

Figure 6: Vertical phase-space distribution upstream of Q021 created directly by G4Beamline and by tomography (bottom). For tomography we use 30 projections.

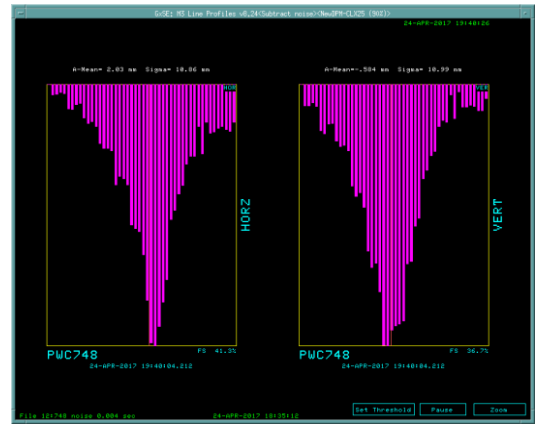

Figure 7: Example of a measured beam profile with a PWC monitor that is placed along the Fermilab Muon Campus.